\documentclass[useAMS,usenatbib,referee]{mn2e}

\usepackage{natbib,graphicx}

\begin{document}

\title[Disruption of the three-body systems]
{Disruption of the three-body gravitational systems: Lifetime
statistics}

\author[V.~V.~Orlov\thanks, A.~V.~Rubinov and I.~I.~Shevchenko]
{V.~V.~Orlov$^1$\thanks{E-mail: vor@astro.spbu.ru},
A.~V.~Rubinov$^1$ and I.~I.~Shevchenko$^2$ \\
$^1$Sobolev Astronomical Institute, St.Petersburg State University,
Universitetsky pr. 28, Stary Peterhoff, St.Petersburg 198504, Russia \\
$^2$Pulkovo Observatory of the Russian Academy of
Sciences, Pulkovskoje ave. 65, St.Petersburg 196140, Russia}

\date{Accepted . Received ; in original form 2010 March 10}


\pubyear{2010}

\maketitle

\label{firstpage}

\begin{abstract}
We investigate statistics of the decay process in the equal-mass
three-body problem with randomized initial conditions. Contrary to
earlier expectations of similarity with ``radioactive decay'', the
lifetime distributions obtained in our numerical experiments turn
out to be heavy-tailed, i.e. the tails are not exponential, but
algebraic. The computed power-law index for the differential
distribution is within the narrow range, approximately from $-1.7$ to
$-1.4$, depending on the virial coefficient. Possible applications
of our results to studies of the dynamics of triple stars known to
be at the edge of disruption are considered.
\end{abstract}


\section{Introduction}

Disruption of a three-body gravitational system is an enigmatic
dynamical process, statistics of which is mostly unexplored, at
long timescales especially. \cite{V88} supposed that the lifetime
distribution for a three-body system is an exponentially decaying
function, in analogy with ``radioactive decay''. Recent
statistical numerical studies by \cite{MT07} of this
process in the equal-mass problem revealed new important data on
the statistics of the disruption times and raised new questions on
the nature of this process. \cite{MT07} have explored a statistics of
the disruption times $T_\mathrm{d}$ in the equal-mass three-body
problem. (The disruption time $T_\mathrm{d}$ is the system
lifetime as a bound system.)
They computed the disruption times for equal-mass systems with
randomized initial conditions, and fitted the found
disruption time distributions by exponential functions.

However, exponential decay, in the asymptotically long time scale,
is in conflict with the theoretical result of \cite{A83} that the
average lifetime of a general isolated three-body system is
infinite. What is more, recently it was shown by \cite{S09} on
the basis of the Kepler map theory, applied to a hierarchical
three-body problem, that at the edge of disruption the orbital
periods and the size of the orbit of the escaping body behave as
L\'evy flights. Due to them, the time decay of the survival
probability (the integral distribution of lifetimes)
is heavy-tailed with the power-law index equal to
$-2/3$. Combining the Kepler map theory and earlier theoretical
findings of \cite{H93}, \cite{S09} made a conclusion that the
$T_\mathrm{d}^{-2/3}$ law is expected to be quite universal.

Here we explore the problem of the lifetime statistics
in the three-body problem by means of
numerical simulations. We pay particular
attention to analysis of the tails of the lifetime distributions,
and find the algebraic behaviour.
We compare our results with previous numerical work
and discuss why the algebraic tails had not been identified
earlier. Possible applications of our results to studies of the
dynamics of triple stars known to be at the edge of disruption are
considered.

\section{Setting of the problem and numerical experiments}
\label{ps}

To investigate disruption of triple systems we use numerical integration of equations
of motion in the gravitational three-body problem.
The equations are as follows:

\begin{equation}
m_k \frac{d^2 {\mathbf r}_k}{dt^2} = \frac{\partial U}{\partial {\mathbf r}_k}
\quad (k=1, 2, 3),
\label{eq:eq_1}
\end{equation}

\noindent where $m_k$ are the masses of the bodies,
${\mathbf r}_k$ are the coordinate vectors in
the barycentric orthogonal coordinate system,
$t$ is time, and $U$ is the potential:

\begin{equation}
U = G \left( \frac{m_1 m_2}{r_{12}}+\frac{m_1 m_3}{r_{13}}+\frac{m_2 m_3}{r_{23}} \right) ,
\label{eq:eq_2}
\end{equation}

\noindent where the quantities $r_{ij}$ are the distances
between bodies $i$ and $j$, $G$ is the gravitational constant.

We use the chain-regularization algorithm of \cite{MA93}
to treat behaviour of the system in the vicinity of close double and triple approaches accurately.
Numerical integration is performed by the Dormand--Prince integrator \citep{HNW87}, which
realizes an explicit Runge--Kutta method of 8-th order with step size control.

We restrict our present study to the equal-mass three-body problem.
The masses of all three bodies $m_k$,
$k = 1,\,2,\,3$ are set to $1$. We also choose $G=1$.
Total energy of the triple system $E=-1$.

To set up the initial conditions for the integration,
we use an approach similar to that used in \cite{MT07}.
Namely, we produce random initial data by selecting the initial
coordinate and velocity components of the bodies
from a uniform distribution
inside a 9-dimensional cube with size of the edge equal to $2$.
Then we transform these initial data
to the barycentric system.

Using scale coefficients, the
coordinates and velocities of bodies are
transformed to satisfy fixed the virial ratio $k = T/U$,
where $T$ and $U$ are the kinetic energy and the
potential, respectively, of the triple system.
The total energy is $E = T-U = -1$.
The following values of the initial virial ratio have been chosen:
$k = 0, 0.1, 0.3, 0.5, 0.7, 0.9$. For each value
of the virial ratio we study a set of $10^5$ triple systems.

We follow the dynamical evolution of each triple system
during the time interval of $10^5$ time units, or until the escape criterion
is satisfied.
The escape criterion is the same as that used in \citep{MT07}; namely,
we stop our integration, when
a single body is moving away on a hyperbolic relative orbit
at a distance $d > 50$ times the current semi-major axis
of the final binary. The condition for hyperbolicity is

\begin{equation}
E_\mathrm{out} = \frac{M_\mathrm{in} V_\mathrm{cm}^2}{2} + \frac{m_\mathrm{out}
V_\mathrm{out}^2}{2} - G \frac{M_\mathrm{in}m_\mathrm{out}}{d} > 0,
\end{equation}

\noindent where $E_\mathrm{out}$ is the total energy of the outer binary,
$M_\mathrm{in}$ is the mass of the inner binary,
$m_\mathrm{out}$ is the mass of the outer component,
$V_\mathrm{cm}$ is the velocity of centre-of-mass of the inner binary,
$V_\mathrm{out}$ is the velocity of the outer component,
$d$ is the distance between centre-of-mass of the inner binary and the outer component.

Using orbital elements we compute the formal time of pericentre passage
for this hyperbolic orbit.
This time $T_\mathrm{d}$ is considered to be the moment of the triple system disruption.
The disruption time $T_\mathrm{d}$ is counted from the beginning of computation
up to this moment.

\section{Results of the numerical experiments}
\label{rne}

The results of integration are qualitatively similar for all
values of $k$ in our set, that is why the graphs with
distributions are presented here only for a single case of $k$,
namely, for $k = 0$.

\begin{figure}
\centering
\includegraphics[width=0.75\textwidth]{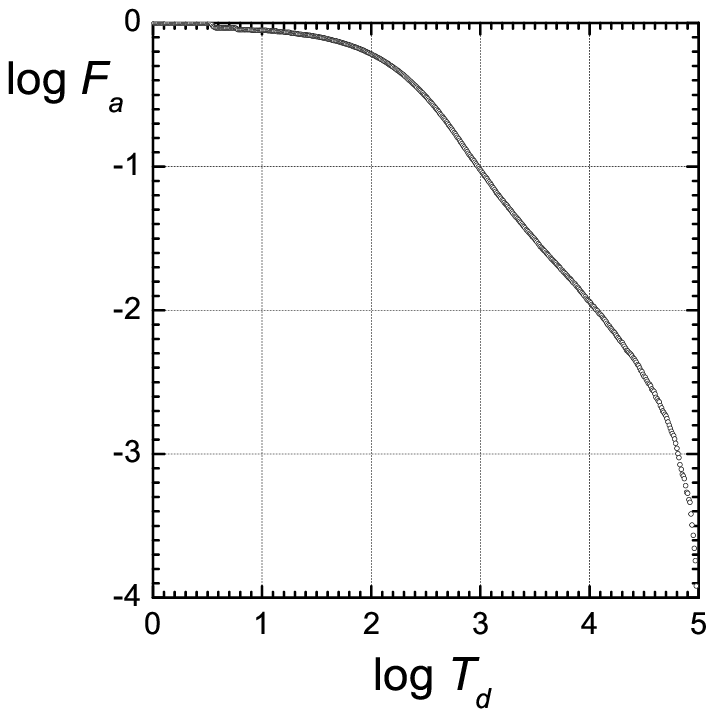}
\caption{The integral distribution of the disruption times in
logarithmic scales, $k = 0$. } \label{log_dd00}
\end{figure}

In Fig.~\ref{log_dd00}, we show the integral distribution
$F_a(T_\mathrm{d})$ of the disruption times in decimal logarithmic
scales. The quantity $F_a(T_\mathrm{d})$ is the fraction of the lifetimes greater than $T_\mathrm{d}$. An initial exponential drop (a ``bump'' in logarithmic
scales) and a subsequent power-law decay (a straight-line
dependence in logarithmic scales) are prominent. In the very tail
a smooth drop is evident, which is due to the presence of the
upper limit on the time of integration (equal to $10^5$ time
units). Owing to this final drop, the $F_a(T_\mathrm{d})$
dependence cannot be used straightforwardly (by linear fitting of
the tail in the logarithmic scales) for reliable estimation of the
index of the power-law decay.

We separate the initial exponential drop from the subsequent
power-law decay by choosing the value of $T_\mathrm{d}$, at which
the distribution becomes algebraic. From Fig.~1 we see that this
transition value lies between $10^3$ and $10^4$. We choose the
latter value, so that to be completely sure that any contribution
of the initial ``bump'' is minimal.

\begin{figure}
\centering
\includegraphics[width=0.75\textwidth]{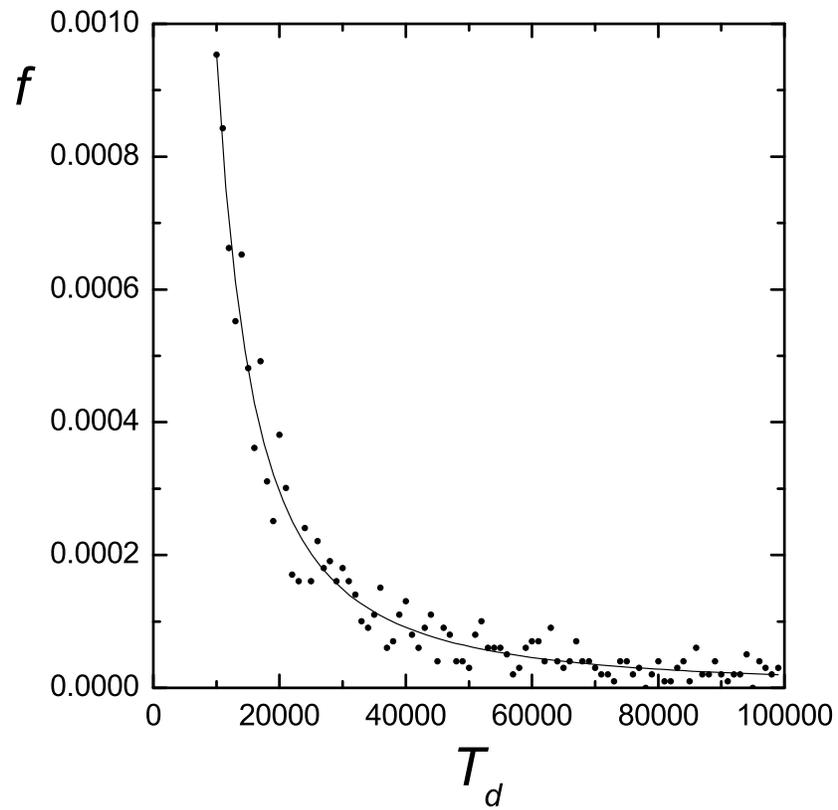}
\caption{The differential distribution of the disruption times for $k = 0$.
The solid line represents the algebraic fitting, which turns
out to be close to $T_\mathrm{d}^{-3/2}$ law.} \label{dd00}
\end{figure}

Thus the transition value of $T_\mathrm{d}$ is found from the
analysis of the bimodal structure of the integral distribution of
$T_\mathrm{d}$ in logarithmic scales. In the subsequent treatment
the scales are no longer logarithmic. In Fig.~\ref{dd00}, the
differential distribution of the disruption times is shown for the
same $k = 0$. Only the tail ($T_\mathrm{d} > 10^4$) is presented.
The quantity $f(T_\mathrm{d})$ is the fraction of the lifetimes
in a pocket ($T_\mathrm{d}$, $T_\mathrm{d} + \Delta$), where
$\Delta = 10^3$. As soon as the distribution is differential, the
upper limit on the time of integration does not inflict the form
of the distribution, and so the distribution can be used for
fitting. The solid line, drawn in the Figure, represents the
algebraic fitting. Details of the fitting are given in Table~1.
For all kinds of fittings in this paper, we use a non-linear
least-squares method \citep{levenberg44,marquardt63} to minimize
$\chi^2$, thereby finding the best-fit parameter values and their
standard errors.

The fitting here is two-parametric: we use the fitting function
\begin{equation}
f(T_\mathrm{d}) = A T_\mathrm{d}^{-\beta},
\end{equation}

\noindent where $A$ and $\beta$ are free parameters. Only the
value of $\beta$ is important for us and is therefore reported in
Table~1. The correlation coefficient $R^2$ is also reported. Note
that it is very close to 1, i.e. the fitting is very good. As one
can see, it turns out to be close to the $T_\mathrm{d}^{-5/3}$
law.

\begin{figure}
\centering
\includegraphics[width=0.75\textwidth]{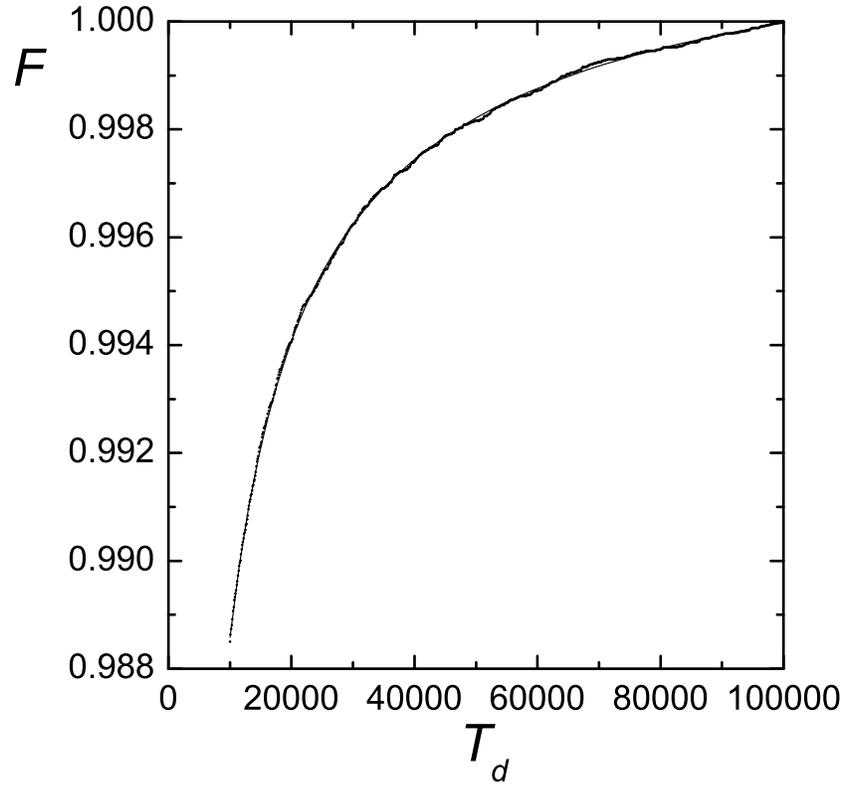}
\caption{The same as in Fig.~\protect{\ref{dd00}}, but the
distribution is integral. The solid line represents the
algebraic fitting.}
\label{id00}
\end{figure}

In Fig.~\ref{id00}, the same data are presented as in
Fig.~\ref{dd00}, but the distribution is integral. The quantity
$F(T_\mathrm{d})$ is the fraction of the lifetimes {\it less}
than $T_\mathrm{d}$. (Note that the quantity $F_a(T_\mathrm{d})$
in Fig.~1 is the fraction of the lifetimes {\it greater} than
$T_\mathrm{d}$). The solid line, drawn in the Figure, represents
the algebraic fitting. A two-parametric fitting cannot be employed
in the integral case, because we do not know beforehand the
fraction of the disruption times longer than the limiting time of
integration (for the differential distribution, this fraction is
absorbed in the coefficient $A$). So, the fitting here is
three-parametric: we use the fitting function
\begin{equation}
F(T_\mathrm{d}) = A - B T_\mathrm{d}^{-\alpha} ,
\end{equation}

\noindent where $A$, $B$ and $\alpha$ are free parameters. Only
the value of $\alpha$ is important for us and is therefore reported
in Table~2.
The fitting turns out to be close to $T_\mathrm{d}^{-2/3}$ law.
This law is expected from our results on the differential
distribution, because the equality $\alpha = \beta - 1$ should
hold. In practice this equality is not exact, mostly due to the
difference in fitting schemes.

The correlation coefficient in the case of the integral
distribution is much closer to 1, than in the case of the
differential distribution. This means that the estimates of
$\alpha$ are more reliable than those of $\beta$. This is natural,
because the fitting results for the differential distribution are
sensitive to the choice of the pocket width.

\begin{table}
\caption{The power law index $\beta$ for the tail of the
differential distribution of disruption times}
\begin{center}
\begin{tabular}[t]{|c|c|c|c|c|c|c|}
\hline
$k$ & $0$ & $0.1$ & $0.3$ & $0.5$ & $0.7$ & $0.9$ \\
\hline
$\beta$ & $1.695$ & $1.639$ & 1.522 & $1.448$ & $1.586$ & $1.684$ \\
& $\pm 0.041$ & $\pm 0.032$ & $\pm 0.031$ & $\pm 0.022$ & $\pm 0.028$ & $\pm 0.057$ \\
\hline
$R^2$  & $0.967$ & $0.977$ & 0.972 & $0.984$ & $0.981$ & $0.935$ \\
\hline
\end{tabular}
\end{center}
\end{table}

\begin{table}
\caption{The power law index $\alpha$ for the tail of the integral
distribution of disruption times}
\begin{center}
\begin{tabular}[t]{|c|c|c|c|c|c|c|}
\hline
$k$ & $0$ & $0.1$ & $0.3$ & $0.5$ & $0.7$ & $0.9$ \\
\hline
$\alpha$ & $0.7073$  & $0.6704$ & 0.4244 & $0.4555$ & $0.6783$ & $0.6332$ \\
& $\pm 0.0016$ & $\pm 0.0013$ & $\pm 0.0022$ & $\pm 0.0018$ & $\pm 0.0014$ & $\pm 0.0041$ \\
\hline
$R^2$  & $0.9998$ & $0.9998$ & $0.9996$ & $0.9997$ & $0.9998$ & $0.9985$ \\
\hline
\end{tabular}
\end{center}
\end{table}

The results of implementing the fitting procedures at all values
of $k$ are collected in Tables~1 and 2. As it can be seen from the
data in Tables~1 and 2, the tails of distributions in all cases
are algebraic. This is our first main result, following from the
fact that the correlation coefficient $R^2$ in all cases is very
close to 1, i.e. the fitting is very good.

Our second main result is that the computed power-law index for
the differential distributions is in the narrow range,
approximately from $-1.7$ to $-1.4$, depending on the virial
coefficient $k$. The consequences are discussed below.

\section{Discussion of the distributions}
\label{discussion}

What is the nature of the observed algebraic decay? One may recall
that a similar algebraic decay was observed by \cite{SS96,SS97} in
numerical experiments on asteroid dynamics. They showed that the
tail of the integral distribution of the time intervals
$T_\mathrm{s}$ between jumps of the orbital eccentricity of
asteroids in the 3/1 mean motion resonance with Jupiter is not
exponential, but algebraic: $F_a \propto T_\mathrm{s}^{-\alpha}$
with $\alpha \sim 1.5$--$1.7$. This decay was due to sticking of
orbits to chaos borders in the phase space of motion. The
stickiness effect determines the character of the distribution of
Poincar\'e recurrences on large timescales: it is algebraic
\citep{CS81,CS84}. Starting with the pioneering work by
\cite{CS81}, the algebraic decay in the recurrence statistics in
Hamiltonian systems with divided phase space was considered, in
particular, in \citep{CS84,C90,C99,CK08}. \cite{C90}, using his
resonant theory of critical phenomena in Hamiltonian dynamics,
justified a value of $3/2$ for the critical exponent $\alpha$.

However, the values of $\alpha$ observed in our numerical
experiments ($\sim 0.4$--0.7) differ radically from those expected
for the stickiness phenomenon ($\sim 1.5$). Therefore, though the
decay is algebraic, it is not related to the stickiness effect.
Then, what is the origin of the algebraic decay here? On the basis
of the Kepler map theory, \cite{S09} considered statistics of the
disruption and Lyapunov times in a hierarchical three-body
problem. It was shown that at the edge of disruption the orbital
periods and the size of the orbit of the escaping body exhibit
L\'evy flights. Due to them, the time decay of the survival
probability is heavy-tailed with the slope power-law index
$\alpha$ equal to $2/3$ (while the relation between the Lyapunov
and disruption times is quasilinear). Combining the Kepler map
theory and earlier theoretical findings of \cite{H93}, \cite{S09}
concluded that the $T_\mathrm{d}^{-2/3}$ law was expected to be
quite universal.

However, neither the $T_\mathrm{d}^{-2/3}$ law, nor even any
other algebraic law, were reported in the numerical
studies by \cite{MT07}. We believe that the reasons are as
follows: solely the initial part of the distribution was
considered by \cite{MT07}. Besides, the algebraic fitting
functions were not used anyway. As noted by \cite{S09}, the
tail of the disruption time distribution in the given problem
should be considered separately from the initial part, because it
corresponds to a different dynamical situation: in the beginning,
the regime of decay might be exponential; see dynamical analogues
in \citep{SS96,SS97,S99} and the theory for the exponential decay
in \citep{C99}.

As follows from Tables~1 and 2, for some values of $k$, namely
intermediate ones $k = 0.3$ and $0.5$, the power-law index
$\beta$ of the differential distribution is equal to
$1.45$--$1.52$, while the power-law index $\alpha$ of the
integral distribution is equal to $0.42$--$0.46$.
Thus the values of $\beta$ and $\alpha$ are close to 3/2 and 1/2,
respectively.
This might testify for the presence of the inverse square root law
$F_a(T_\mathrm{d}) \propto T_\mathrm{d}^{-1/2}$, which is
inherent to free diffusion in the central part of a chaotic layer
until the finite width of the layer becomes
important~\citep[p.~9]{CS81}, see also \citep{S99}.

Concluding, we see that the Kepler map statistics is valid at $k \sim 0$
and $k \sim 1$, while, hypothetically, the free diffusion
seems to dominate at $k \sim 0.5$. As for the case $k \sim 0.5$,
we believe that the free diffusion predominance is merely an effect of
insufficient times of integration, and expect that an extension of
the integration time to greater limits should make apparent the
asymptotic behaviour typical for the L\'evy flights in the
given problem, when
appropriately long timescales are achieved; in other words, we
expect that the asymptotic behaviour with $\alpha \sim 2/3$ is
universal for all values of $k$.

\section{Multiple stars: Dynamics at the edge of stability}
\label{triple_stars}

In application to actual multiple stars,
our results on the statistics of chaotic
decay are appropriate to be used in studies of the dynamics of multiple stars
known nowadays to be near the edge of stability.

A list of such multiple stars comprises:
HD~40887, HD~76644 (ADS~7114$=$ $\iota$~Uma),
HD~217675 ($o$~And), HD~222326 (ADS~16904) \citep{Or05}.
Probably all these systems are quadruple.
However, they contain close binary stars,
therefore in computer simulations we can consider them as triples.
The mass ratios in these systems are:
1)~$0.65:0.52:(0.69+0.2)$;
2)~$0.41:0.42:(1.94+0.82)$;
3)~$4.2:3.4:(6.8+2.3)$;
4)~$1.2:1.3:(0.7+1.0)$
solar masses, respectively.
The ratios of the orbital periods of the outer and inner binaries
are approximately $6,\,21,\,20,\,14$, respectively.
If considered as triples, these systems have a weak hierarchy
\citep{Or05}:
the distant component is not very much far from the inner pair
(with respect to the size of the latter),
and its perturbation on the motion of the latter is not negligible.
The mutual perturbations can be strong enough to modify
the hierarchy and, finally, lead to disruption of the triple.
Thus the property of weak hierarchy implies that
these systems are near the edge of the stability region in the
phase space of the motion; see \citep{Or05} for details.
Though different from the equal-mass case,
the dynamical behaviour of these systems
should be as well dominated by L\'evy flights, if they are close enough
to disruption.
This proximity should be further analyzed and verified.

Of course, a direct observation of L\'evy flights in the dynamics of such systems
would require unrealistic long times of observation. However, when the current
coordinates and velocities of the components are determined observationally
with sufficient accuracy, the L\'evy flights can be studied and analyzed in
numerical simulations of the future dynamical history of the triples, and
estimates of their lifetimes can be obtained.

\section{Conclusions}

We have investigated statistics of the disruption times
$T_\mathrm{d}$ in the equal-mass three-body problem. The
statistics has been explored in massive numerical simulations
with randomized initial conditions. The distributions of the
disruption times have turned out to be heavy-tailed with the
power-law index being within narrow range near the value of $-5/3$
(for the differential distribution). The observed range is from
$-1.7$ to $-1.4$; the value of the slope index slightly depends on the
virial coefficient.

Our result is in conflict with earlier expectations and analogies
with ``radioactive decay'', which implied exponential tails, and
also in conflict with numerical results of
\cite{MT07}, where exponential decay was found. We explain the
divergence with the latter results by the fact that \cite{MT07}
fitted the distribution as a whole (and what is more, taken
at insufficiently long timescale), while the behaviour of the
tails should be treated separately from the initial
exponential drop.

On the other hand, our finding of the algebraic decay is in
agreement with the theoretical result of \cite{A83}, stating that
the average lifetime of a general isolated three-body system is
infinite. Moreover, it is in agreement with the Kepler map
theory \citep{S09}, predicting just the same value of the slope
index ($\sim -5/3$) that we have observed in the majority of our
numerical experiments. This makes evident the existence of L\'evy
flights in the process of decay.

At intermediate values of the virial coefficient, $k \sim 0.5$, we
have observed the slope index $\sim -1.5$ (for the differential
distribution). This is typical for free diffusion in the
central part of a chaotic layer until the finite width of the layer
becomes important. We believe that the free diffusion predominance
is merely an effect of insufficient times of integration. If one
extends the integration time to sufficient limits, the asymptotic
behaviour with the slope index $\sim -5/3$, characteristic of the
L\'evy flights in the given problem, should become apparent.

Several actual multiple stars
are known nowadays to be near the edge of the stability region in the
phase space of the motion, in particular,
HD~40887, HD~76644 (ADS~7114$=$ $\iota$~Uma),
HD~217675 ($o$~And), HD~222326 (ADS~16904) \citep{Or05}.
Numerical integrations, based on initial data from further
high-precision astrometric observations of these systems,
would allow one to reveal long-term evolution features,
in particular, the theoretically predicted L\'evy flights in the
orbital periods and the geometrical sizes of the systems.

\section*{Acknowledgments}

This work was partially supported by the Russian Foundation for
Basic Research (project \# 09-02-00267-a), the Programme of
Fundamental Research of the Russian Academy of Sciences
``Fundamental Problems in Nonlinear Dynamics'',
and the President Programme for Support the Leading Scientific
Schools (project \# 3290.2010.2).


\newpage

\end{document}